\documentstyle[12pt]{article}
\begin{document}
{\pagestyle{empty}
\rightline{TEZU-F-059}
\rightline{EWHA-TH-002}
\rightline{August 1994}
\vskip 1cm
\centerline{\large \bf Relation between Yang-Baxter and Pair
Propagation
Equations }
\centerline{\large \bf in 16-Vertex Models}
\vskip 1cm
\centerline{Changrim Ahn \footnote{E-mail address:
ahn@benz.kotel.co.kr} }
\centerline{{\it Department of Physics} }
\centerline{{\it Ewha Women's University, Seoul 120, Korea} }
\vskip 0.1in
\centerline{Minoru Horibe \footnote{E-mail address:
horibe@newton.apphy.fukui-u.ac.jp}}
\centerline {{\it Department of Physics, Faculty of Education}}
\centerline {{\it Fukui University,Fukui 910, Japan}}
\vskip 0.1in
\centerline{Kazuyasu Shigemoto \footnote{E-mail address:
shigemot@tezukayama-u.ac.jp}}
\centerline {{\it Department of Physics}}
\centerline {{\it Tezukayama University, Nara 631, Japan }}
\vskip 1cm
\vskip 1cm
\centerline{\bf Abstract} \vspace{10mm}
%\vskip 0.2in
We study a relation between two integrability conditions,
namely the Yang-Baxter and the pair propagation equations, in 2D lattice
models. While the two are equivalent in the 8-vertex
models, discrepancies appear in the 16-vertex models.
As explicit examples, we find the
exactly solvable 16-vertex models which do not satisfy
the Yang-Baxter equations.
%PACS number(s):
\newpage}
%%%%%%%%%%%%%%%%%%%%%%% Section 1 %%%%%%%%%%%%%%%%%%%%%%%%
\noindent {\bf 1.\ \ Introduction} \vspace{3mm}\\
\indent
In last two decades, much progress has been made in 2D integrable
systems both in lattice statistical models and in continuum field theories.
Recently, this progress has been associated with beautiful
mathematical structures such as
universal Grassmann manifold~\cite{Grass}, Kac-Moody
algebra~\cite{Kac} and quantum group~\cite{qg}.
\hfil\break\indent
In 2D lattice models, there is one approach, which is based
on transfer matrices (TMs) and it
has been proved most successful.
As Baxter showed, one can construct infinite number of commuting conserved
quantities through these TMs~\cite{Baxter}.
A sufficient condition for the commutativity is that the Boltzmann weights
of the 2D lattice models satisfy the famous Yang-Baxter equations (YBEs).
There can obviously exist many exactly solvable models which do not satisfy
YBEs. Since these are exactly solvable, one needs another scheme to solve
these models if they exist.
\hfil\break
\indent
There is another approach, which is based on the so-called
pair propagation equations (PPEs)
appearing in the analysis of the algebraic Bethe ansatz.
According to this method, the Boltzmann weights satisfy non-linear
coupled equations. These equations become manageable if the Boltzmann
weights are defined on some algebraic curves.
\hfil\break
\indent
In this paper, we want to study some 2D lattice models which can be exactly
solvable while they do not satisfy YBEs.
We are looking for our candidates from the 16-vertex
models~\cite{Feld}\cite{Bellon}.
What we are going to show first is a relationship between YBEs and PPEs.
Though YBEs and PPEs are equivalent in the 8-vertex model,
discrepancies appear in 16-vertex models.
Since YBEs restrict possible candidates so strongly,
PPEs can cover more exactly solvable models which do not satisfy YBEs.
We give explicit examples for which we compute exact eigenvalues of
transfer matrices. \vspace{10mm}\\
%%%%%%%%%%%%%%%%%%%%%%%%%%% section 2 %%%%%%%%%%%%%%%%%%%%%%
\noindent {\bf 2. The Pair Propagation and Conjugate Pair
Propagation Equations}\vspace{3mm}\\
\indent
We follow the notation of Baxter~\cite{Baxter}.
The Boltzmann weights of the symmetric 16-vertex models are given by
\begin{eqnarray}
R(\pm,\pm;\pm,\pm)=a, R(\pm,\mp;\pm,\mp)=b,
R(\pm,\mp;\mp,\pm)=c,R(\pm,\pm;\mp,\mp)=d,
\nonumber \\
R(\pm,\mp;\mp,\mp)=e,R(\pm,\pm;\mp,\pm)=k,
R(\pm,\pm;\pm,\mp)=h,R(\mp,\pm;\mp,\mp)=l.
\label{e1}
\end{eqnarray}
\indent
The Yang-Baxter equations are given in the forms;
\begin{eqnarray}
\sum_{\eta, \zeta, \phi} R(\mu,\zeta;\eta,\beta)
R^{'}(\rho,\alpha;\phi,\zeta )R^{''}(\eta,\phi;\nu,\sigma)
           \nonumber  \\
=\sum_{\eta, \zeta, \phi} R^{''}(\mu,\rho;\eta,\phi)
R^{'}(\phi,\zeta;\sigma,\beta) R(\eta,\alpha;\nu,\zeta).
                                \label{e8}
\end{eqnarray}
\indent
According to the Bethe ansatz, eigenfunctions
$y(\beta_1,\beta_2,...,\beta_N)$
of transfer matrices $T(v)$ for $N$ horizontal sites become in the
forms of the direct products of each variables such as
$y(\beta_1,\beta_2,...,\beta_N)=g_1(\beta_1)
\otimes g_2(\beta_2)\otimes...\otimes g_N(\beta_N)$. These are
eigenfunctions of transfer matrix on the upper layer.
We multiply these eigenfunctions to transfer matrices, and
obtain
\begin{equation}
\left(T(v)y\right)_{\alpha}={\rm Tr}\left(G_1(\alpha_1)...G_N(\alpha_N)
\right),\quad {\rm with}\quad
\left( G_i(\alpha)\right)_{\mu \nu}
=\sum_{\beta}R_(\mu,\alpha;\nu,\beta) g_i(\beta).   \label{e24}
\end{equation}
Explicit forms of $G_i(\pm)$ are
\begin{eqnarray}
G_i(+)=
\left(\begin{array}{cc}
ag_i(+)+hg_i(-) & kg_i(+)+dg_i(-)  \\
eg_i(+)+cg_i(-) & bg_i(+)+lg_i(-)
\end{array} \right) ,           \nonumber  \\
G_i(-)=
\left( \begin{array}{cc}
lg_i(+)+bg_i(-) & cg_i(+)+eg_i(-)  \\
dg_i(+)+kg_i(-) & hg_i(+)+ag_i(-)
\end{array} \right).     \nonumber
\end{eqnarray}
In order to be solved exactly, it is necessary that there exist
the $\alpha$-independent pairs of matrices $P_i,P_{i+1}$, which
transform $G_i(\alpha)$ into upper triangle forms;
\begin{equation}
P_i^{-1} G_i(\alpha) P_{i+1} =H_i(\alpha)=
\left( \begin{array}{cc}
g_i^{'}(\alpha) & g_i^{'''}(\alpha)  \\
0          & g_i{''}(\alpha)
\end{array} \right).      \label{e27}
\end{equation}
For simplicity, we choose $\det P_i=1$ for $i=1,2,...,N$
and we parametrize them in the forms;
$P_i=
\left( \begin{array}{cc}
p_i(+) & t_i(+)   \\
p_i(-) & t_i(-)
\end{array} \right)$. \
Then Eq.~(\ref{e27}) is written in the forms ;
\begin{equation}
G_i(\alpha) P_{i+1}=P_i H_i(\alpha)
,\quad {\rm or}\quad
P_{i}^{-1} G_i(\alpha)=H_i(\alpha) P_{i+1}^{-1}.    \label{e30}
\end{equation}
As $H_i(\alpha)$ are in upper triangle forms, we obtain;
\begin{eqnarray}
&&G_i(\alpha)
\left(\begin{array}{c}
p_{i+1}(+)   \\
p_{i+1}(-)
\end{array} \right)
= g_{i}^{'}(\alpha)
\left(\begin{array}{c}
p_{i}(+)   \\
p_{i}(-)
\end{array} \right),   \label{e31}  \\
&&\left(\begin{array}{cc}
-p_{i}(-), &  p_{i}(+)
\end{array} \right)
G_i(\alpha)
= g_{i}^{''}(\alpha)
\left(\begin{array}{cc}
-p_{i+1}(-), &  p_{i+1}(+)
\end{array} \right) .  \label{e32}
\end{eqnarray}
We call Eq.~(\ref{e31}) as the pair propagation equations (I)
and Eq.~(\ref{e32}) as the conjugate pair
propagation equations (I).
By using $R(\mu,\alpha;\nu,\beta)$, the pair propagation
equations (I) are given by
\begin{equation}
\sum_{\beta,\nu} R(\mu,\alpha;\nu,\beta) g_i(\beta) p_{i+1}(\nu)
=g_{i}^{'}(\alpha) p_i(\mu)   .           \label{e33}
\end{equation}
Explicit forms of these equations are
\begin{eqnarray}
\left\{ \begin{array}{l}
(ag_i(+)+hg_i(-))p_{i+1}(+)+(kg_i(+)+dg_i(-))p_{i+1}(-)
=g_{i}^{'}(+)p_i(+),               \\
(lg_i(+)+bg_i(-))p_{i+1}(+)+(cg_i(+)+eg_i(-))p_{i+1}(-)
=g_{i}^{'}(-)p_i(+),             \\
(eg_i(+)+cg_i(-))p_{i+1}(+)+(bg_i(+)+lg_i(-))p_{i+1}(-)
=g_{i}^{'}(+)p_i(-),              \\
(dg_i(+)+kg_i(-))p_{i+1}(+)+(hg_i(+)+ag_i(-))p_{i+1}(-)
=g_{i}^{'}(-)p_i(-).  \end{array}\right.   \label{e34}
\end{eqnarray}
While the conjugate pair propagation equations (I) are given by
\begin{equation}
\sum_{\beta,\mu} R(\mu,\alpha;\nu,\beta) g_i(\beta) q_{i}(\mu)
=g_{i}^{''}(\alpha) q_{i+1}(\nu) , \label{e35}
\end{equation}
where we use notation $q_i(+)=-p_i(-),q_i(-)=p_i(+)$.
Explicit forms of these equations are obtained from
Eq.~(\ref{e34}) by replacing
$g_{i}^{'}(\pm)\rightarrow g_{i}^{''}(\pm)$, $p_{i+1}(+)\rightarrow
-p_{i}(-)$, $p_{i+1}(-)\rightarrow p_{i}(+)$,
$p_{i}(+)\rightarrow -p_{i+1}(-)$, $p_{i}(-)\rightarrow p_{i+1}(+)$,
$c \leftrightarrow d$, $e \leftrightarrow k$.
\hfil\break
\indent
Next we consider the second type of the pair propagation equations,
which we call the pair propagation equations (II).
If the models are exactly solvable by using eigenfunctions
$y(\beta_1,\beta_2,...,\beta_N)=g_1(\beta_1)
\otimes g_2(\beta_2)\otimes...\otimes g_N(\beta_N)$
acting on upper layer of the transfer matrices,
it is exacly solvable by using eigenfunction
$\tilde{y}(\alpha_1,\alpha_2,...,\alpha_N)
=\tilde{g}_1(\alpha_1)\otimes \tilde{g}_2(\alpha_2)
\otimes...\otimes \tilde{g}_N(\alpha_N)$ acting on
lower layer. Similar equations corresponding to
Eqs.~(\ref{e24}),(\ref{e27}) are given by
\begin{equation}
\left(\tilde{y}T(v)\right)_{\beta}={\rm Tr}
\left(\tilde{G}_1(\beta_1)
...\tilde{G}_N(\beta_N)\right)
,\quad{\rm with}\quad
\left( \tilde{G}_i(\beta)\right)_{\mu \nu}
=\sum_{\alpha}R_(\mu,\alpha;\nu,\beta) \tilde{g}_i(\alpha),
                                                    \label{e38}
\end{equation}
and
\begin{eqnarray}
\tilde{P}_i^{-1} \tilde{G}_i(\beta) \tilde{P}_{i+1}
=\tilde{H}_i(\alpha)=
\left( \begin{array}{cc}
\tilde{g}_i^{'}(\beta) & \tilde{g}_i^{'''}(\beta)  \\
0          & \tilde{g}_i{''}(\beta)
\end{array} \right).      \label{e39}
\end{eqnarray}
The pair propagation equations (II) are given by
\begin{equation}
\sum_{\alpha,\nu} R(\mu,\alpha;\nu,\beta)
\tilde{g}_i(\alpha) \tilde{p}_{i+1}(\nu)
=\tilde{g}_{i}^{'}(\beta) \tilde{p}_i(\mu).  \label{e40}
\end{equation}
By the symmetry of the Boltzmann weights, explicit forms of
these pair propagation equations (II) are obtained from the pair
propagation equations (I) by replacing
untilde variables into tilde variables and
$c\leftrightarrow d, h\leftrightarrow l$.
Similarly we obtain the conjugate pair propagation equations (II)
from Eq.~(\ref{e35}) by the same replacement.
\hfil\break
\indent
Our strategy to solve the pair propagation equations is the
following.
These pair propagation and conjugate pair propagation
equations are special bilinear equations of four variables
such as $g_{i}(\pm),\ g_{i}^{'}(\pm), p_{i}(\pm),p_{i+1}(\pm)$,
and it is difficult to solve these equations directly.
Then we first derive non-linear equations,
where only two ratio of the variables
such as $r_{i}=p_{i}(-)/p_{i}(+),\ r_{i+1}=p_{i+1}(-)/p_{i+1}(+)$
 appear. Instead of solving four coupled equations in the pair
propagation equations, we first solve these four
non-linear equations. For each solution of these
non-linear but two variable equations, only one of four equations of
the pair propagation equations is independent, and from that we
can obtain the eigenvalues of the transfer matrices.
\hfil\break
\indent
{}From  condition that there exists non-trivial solutions for
$g_{i}(\pm),g_{i}^{'}(\pm)$\ in Eq.~(\ref{e34}), we obtain
\begin{eqnarray}
&&r_{i}^2+r_{i+1}^2-\Gamma_1 (r_{i}^2 r_{i+1}^2+1)-\Gamma_2 r_{i}
r_{i+1} +\Gamma_3 r_{i}(1+r_{i+1}^2)+\Gamma_4 r_{i+1}(1+r_{i}^2)=0
\label{e43}  \\
&&{\rm where}   \nonumber  \\
&&\left\{ \begin{array}{l}
\Gamma_1=(cd-ek)/(ab-hl), \quad
\Gamma_2=(a^2+b^2+e^2+k^2-c^2-d^2-h^2-l^2)/(ab-hl), \\
\Gamma_3=(cl+dh-ak-be)/(ab-hl), \quad
\Gamma_4=(ae+bk-ch-dl)/(ab-hl)   \end{array} \right.
\nonumber \\
&&{\rm and}   \nonumber \\
&&r_{i}=p_{i}(-)/p_{i}(+),
\quad r_{i+1}=p_{i+1}(-)/p_{i+1}(+) . \nonumber
\end{eqnarray}
\indent
{}From the condition that there exists non-trivial
solutions for $p_{i}(\pm),p_{i+1}(\pm)$, we obtain
\begin{eqnarray}
&&s_{i}^2+{s_{i}^{'}}^2-\Gamma_{5} (s_{i}^2 {s_{i}^{'}}^2+1)
-\Gamma_{6} s_{i} s_{i}^{'}
+\Gamma_{7} s_{i}^{'}(1+s_{i}^2)
+\Gamma_{8} s_{i}(1+{s_{i}^{'}}^2)=0     \label{e45}  \\
&&{\rm where}   \nonumber  \\
&&\left\{ \begin{array}{l}
\Gamma_{5}=(cd-hl)/(ab-ek), \quad
\Gamma_{6}=(a^2+b^2+h^2+l^2-c^2-d^2-e^2-k^2)/(ab-ek), \\
\Gamma_{7}=(ce+dk-ah-bl)/(ab-ek), \quad
\Gamma_{8}=(al+bh-ck-de)/(ab-ek),   \end{array} \right.
\nonumber \\
&& {\rm and}   \nonumber \\
&& s_{i}=g_{i}(-)/g_{i}(+),
\quad s^{'}_{i}=g_{i}^{'}(-)/g_{i}^{'}(+).  \nonumber
\end{eqnarray}
\indent
Simlarly, from the condition that there exists non-trivial
solutions for $g_{i}^{'}(\pm),p_{i+1}(\pm)$,we obtain
\begin{eqnarray}
&&r_{i}^2+{s_{i}}^2-\Gamma_{9} (r_{i}^2 {s_{i}}^2+1)
-\Gamma_{10} r_{i} s_{i}
+\Gamma_{11} r_{i}(1+s_{i}^2)
+\Gamma_{12} s_{i}(1+r_{i}^2)=0        \label{e47}  \\
&&{\rm where}   \nonumber  \\
&&\left\{\begin{array}{l}
\Gamma_{9}=(bd-eh)/(ac-kl), \quad
\Gamma_{10}=(a^2+c^2+e^2+h^2-b^2-d^2-k^2-l^2)/(ac-kl),   \\
\Gamma_{11}=(bl+dk-ah-ce)/(ac-kl), \quad
\Gamma_{12}=(ae+ch-bk-dl)/(ac-kl).    \end{array} \right.
\nonumber
\end{eqnarray}
\indent
Finally, from the condition that there exists non-trivial
solutions for $g_{i}(\pm),p_{i}(\pm)$, we obtain
\begin{eqnarray}
&&{s_{i}^{'}}^2+r_{i+1}^2-\Gamma_{13} ({s_{i}^{'}}^2r_{i}^2+1)
-\Gamma_{14} s_{i}^{'} r_{i+1}
+\Gamma_{15} s_{i}^{'}(1+r_{i+1}^2)
+\Gamma_{16} r_{i+1}(1+{s_{i}^{'}}^2)=0    \label{e49}  \\
&&{\rm where}   \nonumber  \\
&&\left\{ \begin{array}{l}
\Gamma_{13}=(bd-kl)/(ac-eh), \quad
\Gamma_{14}=(a^2+c^2+k^2+l^2-b^2-d^2-e^2-h^2)/(ac-eh),  \\
\Gamma_{15}=(be+dh-ak-cl)/(ac-eh), \quad
\Gamma_{16}=(al+ck-bh-de)/(ac-eh).  \end{array} \right.
\nonumber
\end{eqnarray}
\indent
While we obtain the
conjugate pair propagation equations (II) from the pair propagation
equations (I)
by replacing $p_i,p_{i+1},g_i,g_{i}^{'}
\rightarrow \tilde{p}_i,\tilde{p}_{i+1},\tilde{g}_i,\tilde{g}_{i}^{'}$
, that is, by replacing ratios $r_i,r_{i+1},s_{i},s_{i}^{'} \rightarrow
\tilde{r}_{i},\tilde{r}_{i+1},\tilde{s}_{i},\tilde{s}_{i}^{'}$
and $c \leftrightarrow d$, $h \leftrightarrow l$.
Explicit forms are given by
\begin{eqnarray}
&&\tilde{r}_{i}^2+\tilde{r}_{i+1}^2
-\Gamma_1 (\tilde{r}_{i}^2 \tilde{r}_{i+1}^2+1)
-\Gamma_2 \tilde{r}_{i} \tilde{r}_{i+1}
+\Gamma_3 \tilde{r}_{i}(1+\tilde{r}_{i+1}^2)
+\Gamma_4 \tilde{r}_{i+1}(1+\tilde{r}_{i}^2)=0  \label{e50}  \\
&&\tilde{s}_{i}^2+{\tilde{s_{i}^{'}}}^2
-\Gamma_{5} (\tilde{s}_{i}^2 {\tilde{s_{i}^{'}}}^2+1)
-\Gamma_{6} \tilde{s}_{i} \tilde{s_{i}}^{'}
-\Gamma_{8} \tilde{s}_{i}^{'}(1+\tilde{s}_{i}^2)
-\Gamma_{7} \tilde{s}_{i}(1+{\tilde{s_{i}^{'}}}^2)=0 .
\label{e52}
\end{eqnarray}\vspace{7mm}\\
%%%%%%%%%%%%%%%%%%%%%%%%%%% section 3 %%%%%%%%%%%%%%%%%%%%%%
\noindent {\bf 3. Connection between the Yang-Baxter and
the Pair Propagation Equations}\vspace{3mm}\\
\indent
Next we consider the connection between the Yang-Baxter and the pair
propagation equations.  We consider products of three $R$-matrices
in Eq.~(\ref{e8}) as  matrices with row indexed by $\beta,\mu,\rho$
and with column indexed by $\alpha,\nu,\sigma$. We denote quantities in
left-hand side as $A(\beta,\mu,\rho|\alpha,\nu,\sigma)$ and those
in right-hand side as
$B(\beta,\mu,\rho|\alpha,\nu,\sigma)$.
To show the relations
$A(\beta,\mu,\rho|\alpha,\nu,\sigma)
=B(\beta,\mu,\rho|\alpha,\nu,\sigma)$ is equivalent to show
\begin{equation}
\sum_{\alpha,\nu,\rho} A(\beta,\mu,\rho|\alpha,\nu,\sigma)
v_{1}(\alpha)v_{2}(\nu)v_{3}(\sigma)
=\sum_{\alpha,\nu,\rho} B(\beta,\mu,\rho|\alpha,\nu,\sigma)
v_{1}(\alpha)v_{2}(\nu)v_{3}(\sigma).
\label{e42}
\end{equation}
for arbitrary three vectors
$v_{1}(\alpha)v_{2}(\nu)v_{3}(\sigma)$.
Explicit forms of these equations are given by
\begin{eqnarray}
\sum_{\eta,\zeta,\phi,\alpha,\nu,\sigma} R(\mu,\zeta;\eta,\beta)
R^{'}(\rho,\alpha;\phi,\zeta)R^{''}(\eta,\phi;\nu,\sigma)
v_1(\alpha)v_2(\nu)v_3(\sigma)
           \nonumber  \\
=\sum_{\eta, \zeta, \phi,\alpha,\nu,\sigma} R^{''}(\mu,\rho;\eta,\phi)
R^{'}(\phi,\zeta;\sigma,\beta)R(\eta,\alpha;\nu,\zeta)
v_1(\alpha)v_2(\nu)v_3(\sigma) ,                  \label{e53}
\end{eqnarray}
We can transform these by using the pair propagation equations as
\begin{eqnarray}
{\rm (left-hand\ side)}
&&\equiv \sum_{\eta, \zeta, \phi,\alpha} R(\mu,\zeta;\eta,\beta)
R^{'}(\rho,\alpha;\phi,\zeta)
v_1(\alpha)u_2^{'}(\eta)u_3^{'}(\phi) \nonumber \\
&&\equiv \sum_{\eta,\zeta}
R(\mu,\zeta;\eta,\beta)
t_1^{'}(\zeta)u_2^{'}(\eta)t_3^{'}(\rho) \nonumber \\
&&\equiv z_1{'}(\beta) z_2{'}(\mu) t_3{'}(\rho) ,  \label{e54}  \\
{\rm (right-hand\ side)}
&&\equiv \sum_{\eta, \zeta, \phi,\sigma} R^{''}(\mu,\rho;\eta,\phi)
R^{'}(\phi,\zeta;\sigma,\beta)
u_1{''}(\zeta)u_2{''}(\eta)v_3(\sigma) \nonumber  \\
&&\equiv \sum_{\eta, \phi} R^{''}(\mu,\rho;\eta,\phi)
t_1{''}(\beta)u_2{''}(\eta)t_3{''}(\phi) \nonumber \\
&&\equiv t_1{''}(\beta)z_2{''}(\mu)z_3{''}(\rho).     \label{e55}
\end{eqnarray}
Ratios of vectors change in the following way;
\begin{eqnarray}
{\rm (left-hand\ side)}:
\left\{ \begin{array}{c}
v_1(-)/v_1(+) \quad \begin{picture}(20,20)(0,0) \put(0,0)
{$\longrightarrow$} \put(2,-10){(II)} \end{picture}\quad
t_1^{'}(-)/t_1^{'}(+) \quad \begin{picture}(20,20)(0,0) \put(0,0)
{$\longrightarrow$} \put(3,-10){(I)} \end{picture}\quad
z_1^{'}(-)/z_1^{'}(+),     \\
v_2(-)/v_2(+) \quad \begin{picture}(20,20)(0,0) \put(0,0)
{$\longrightarrow$} \put(3,-10){(I)} \end{picture}\quad
u_2^{'}(-)/u_2^{'}(+) \quad \begin{picture}(20,20)(0,0)
\put(0,0) {$\longrightarrow$} \put(3,-10){(I)} \end{picture}\quad
z_2^{'}(-)/z_2^{'}(+),     \\
v_3(-)/v_3(+) \quad \begin{picture}(20,20)(0,0) \put(0,0)
{$\longrightarrow$} \put(1,-10){(III)} \end{picture}\quad
u_3^{'}(-)/u_3^{'}(+) \quad \begin{picture}(20,20)(0,0)
\put(0,0) {$\longrightarrow$} \put(1,-10){(III)}
\end{picture}\quad t_3^{'}(-)/t_3^{'}(+),
\end{array} \right.
                                   \label{e56}
\end{eqnarray}
\begin{eqnarray}
{\rm (right-hand\ side)}:
\left\{ \begin{array}{c}
v_1(-)/v_1(+) \quad \begin{picture}(20,20)(0,0) \put(0,0)
{$\longrightarrow$} \put(3,-10){(I)} \end{picture}
\quad  u_1^{''}(-)/u_1^{''}(+) \quad
\begin{picture}(20,20)(0,0) \put(0,0)
{$\longrightarrow$} \put(2,-10){(II)} \end{picture} \quad
t_1^{''}(-)/t_1^{''}(+),          \\
v_2^{''}(-)/v_2^{''}(+) \quad \begin{picture}(20,20)(0,0)
\put(0,0) {$\longrightarrow$} \put(3,-10){(I)}
\end{picture}\quad   u_2^{''}(-)/u_2^{''}(+)
\quad \begin{picture}(20,20)(0,0) \put(0,0)
{$\longrightarrow$} \put(3,-10){(I)} \end{picture}\quad
z_2^{''}(-)/z_2^{''}(+),          \\
v_3^{''}(-)/v_3^{''}(+) \quad \begin{picture}(20,20)(0,0)
\put(0,0) {$\longrightarrow$} \put(1,-10){(III)}
\end{picture}\quad   t_3^{''}(-)/t_3^{''}(+) \quad
\begin{picture}(20,20)(0,0) \put(0,0) {$\longrightarrow$}
\put(1,-10){(III)} \end{picture}\quad
z_3^{''}(-)/z_3^{''}(+),
\end{array} \right.
                                   \label{e57}
\end{eqnarray}
where we use equations (I), (II) and (III), which
connect {\it in} variable $X$
with {\it out} variable $Y$
in the following forms;
\begin{eqnarray}
\left\{ \begin{array}{l}
{\rm (I)}:X^2+Y^2-\Gamma_1(X^2 Y^2 +1)
-\Gamma_2 X Y
+\Gamma_3 Y (1+X^2)
+\Gamma_4 X (1+Y^2)=0,            \\
{\rm (II)}:X^2+Y^2-\Gamma_{5}(X^2 Y^2 +1)
-\Gamma_{6} X Y
+\Gamma_{7} Y (1+X^2)
+\Gamma_{8} X (1+Y^2)=0,        \\
{\rm (III)}:X^2+Y^2-\Gamma_{5}(X^2 Y^2 +1)
-\Gamma_{6} X Y
-\Gamma_{8} Y (1+X^2)
-\Gamma_{7} X (1+Y^2)=0.  \end{array} \right. \nonumber
\end{eqnarray}
If the forms of Eqs.~(I) and (II) are the same,
Eq.~(\ref{e53}) is satisfied.
Conditions that the forms of Eqs.~(I) and (II) are the same lead
$\Gamma_{i}=\Gamma_{i+4}, (i=1 \sim 4)$.
We call these conditions as the candidate
conditions to satisfy the Yang-Baxter equations,
because these conditions do not
guarantee to satisfy the Yang-Baxter equations but mean to satisfy
three vectors multiplied forms of the Yang-Baxter equations.
\hfil\break
\indent
In the 16-vertex model case, the above candidate conditions lead further
restrictions on the Boltzmann weights, that is, we obtain two
possibilities $i)\ a+d=b+c, \ e=h, \ k=l \ $ \ and $ii)\ e=l, \ k=h \ $.
Taking into account of these considerations to find the candidates,
we consider the following exactly solvable cases, which are the more
restricted cases than the above possibilities,
$i)a=c,\ b=d,\ e=h,\ k=l$\ and $ii)a=d,\ b=c,\ e=l,\ k=h$ \ in
the next section.
\vspace{10mm}\\

%%%%%%%%%%%%%%%%%%%%%%%%%%% section 4 %%%%%%%%%%%%%%%%%%%%%%
{\bf 4. Exactly Solvable Cases in 16-Vertex Models}

\indent
The whole regime of 16 vertex model is not yet
solved exactly, and further, not yet shown to be integrable.
In this situation, we suspect
that the 16 vertex model will be not integrable in the whole regime.
Then in the following, we will restrict the original model to more
specialized cases, expecting to find integrable cases.
This means that we will
not cover the whole regime of the original model,
but will cover the entire temperature range as we will find
the whole spectrum of the transfer matrices.
Furthermore, as our main poit of our paper is the relation
between Yang-Baxter and pair propagation equations,
we will examine some special cases of the original model
in order to demonstrate the method of
section 3, but we will not intend to analyze the whole regime
completely. \vspace{3 mm}

{\it i) $a=c, b=d, e=h, k=l$ case }\\
\indent
In this case, we find that the Yang-Baxter equations are
satisfied in a sense that
there always exists a non-trivial set
$\{a{''},\ ,b^{''},\ ,e^{''},\ k^{''}\}$\ for given sets
$\{a,\ ,b,\ ,e,\ k\},\ \{a^{'},\ ,b^{'},\ ,e^{'},\ k^{'}\}$.
By explicit calculations, apparent independent Yang-Baxter
equations reduce from $(32-4)=28$\ to $3$ in the forms;
\begin{eqnarray}  \left\{ \begin{array}{l}
e^{''}C_{\alpha}-k^{''}C_{\beta}=0      \\
(a^{''}-b^{''})C_{\beta}-e^{''}C_{\delta}=0    \\
a^{''}C_{\xi}-b^{''}C_{\eta}=0   \end{array} \right. \label{e74}
\end{eqnarray}
where $C_{\alpha}=ak{'}+a^{'}k+be^{'}+b^{'}e,\
C_{\beta}=ae{'}+a^{'}e+bk^{'}+b^{'}k,\
C_{\delta}=(a-b)(a^{'}-b^{'})+(e-k)(e^{'}-k^{'}),\
C_{\eta}=aa{'}+bb^{'}+ee^{'}+kk^{'},\
C_{\xi}=ab{'}+a^{'}b+ek^{'}+e^{'}k $.
Then for given sets
$\{a,\ ,b,\ ,e,\ k\},\ \{a^{'},\ ,b^{'},\ ,e^{'},\ k^{'}\}$,
there always exists a non-trivial set
$\{a{''},\ ,b^{''},\ ,e^{''},\ k^{''}\}$, that is, the Yang-Baxter equations
are always satisfied.
\hfil\break
\indent
In this case, Eqs.~(\ref{e43}), (\ref{e45}), (\ref{e47}), (\ref{e49})
give
\begin{eqnarray} \left\{ \begin{array}{l}
 (r_{i}^2-1)(r_{i+1}^2-1)=0,      \\
 (s_{i}^2-1)({s_{i}^{'}}^2-1)=0,      \\
 r_{i}^2+s_{i}^2-\Gamma_{9}({r_{i}}^2 {s_{i}}^2 +1)
-\Gamma_{10}r_{i}s_{i}
+\Gamma_{11}\left(r_{i}(1+s_{i}^2)-s_{i}(1+r_{i}^2)\right)
=0,                           \\
{s_{i}^{'}}^2+r_{i+1}^2-\Gamma_{13}({s_{i}^{'}}^2 {r_{i+1}}^2 +1)
-\Gamma_{14}r_{i}s_{i}
+\Gamma_{15}\left(s_{i}^{'}(1+r_{i+1}^2)
-r_{i+1}(1+{s_{i}^{'}}^2)\right) =0,  \end{array} \right. \label{e80}
\end{eqnarray}
where
%\hfil\break
\begin{eqnarray} \left\{ \begin{array}{l}
\Gamma_9=(b^2-e^2)/(a^2-k^2),
\Gamma_{10}=2(a^2+e^2-b^2-k^2)/(a^2-k^2),
\Gamma_{11}=2(bk-ae)/(a^2-k^2),\\
\Gamma_{13}=(b^2-k^2)/(a^2-e^2),
\Gamma_{14}=2(a^2+k^2-b^2-e^2)/(a^2-e^2),
\Gamma_{15}=2(be-ak)/(a^2-e^2). \end{array} \right.
\nonumber
\end{eqnarray}
Solutions are the combinations of $r_{i}=s_{i}=\pm 1$\ and
$r_{i+1}=s_{i}^{'}=\pm 1$, (signs are independent for both).
Then we obtain following cases;
\hfil\break
{\it ia) $r_i=s_i=r_{i+1}=s_{i}^{'}=\pm 1$ case}
\hfil\break
\indent
Choosing $p_{i}=p_{i+1}$\ , eigenfunctions at this site become
\begin{equation}
g_{i}=
\left(\begin{array}{c}
g_{i}(+)  \\  \pm g_{i}(+) \\ \end{array}\right)
,\quad
g_{i}^{'}
=(a+b \pm(e+k)) g_{i}
,\quad
g_{i}^{''}=0.                        \nonumber
\end{equation}
{\it ib) $r_i=s_i=-r_{i+1}=-s_{i}^{'}=\pm 1$ case}
\hfil\break
\indent
Choosing $p_{i}=p_{i+1}$\ , eigenfunctions become
\begin{equation}
g_{i}=
\left(\begin{array}{c}
g_{i}(+)  \\ \pm g_{i}(+) \\ \end{array}\right) ,\quad   \nonumber \\
g_{i}^{'}
=(a-b \pm(e-k))
\left(\begin{array}{c}
g_{i}(+)  \\  \mp g_{i}(+) \\ \end{array}\right) ,\quad
g_{i}^{''}=0.   \label{e83}
\end{equation}
When we mix these cases, it is necessary to take $r_{1}=r_{N+1}$\ for
periodicity, but otherwise we can mix {\it ia)}\ and {\it ib)}\ in such a way
as (the number of times of mixing $r_i=-r_{i+1}=1$\ cases)$=$
(the number of times of mixing $r_i=-r_{i+1}=-1$\ cases).  General
eigenvalues of transfer matrices give
\begin{eqnarray}
\Lambda=(a+b+e+k)^{m_{1}}(a+b-e-k)^{m_{2}}((a-b)^2-(e-k)^2)^{m_{3}}
(\pm 1)^{m_{3}},                 \label{e85}
\end{eqnarray}
by using non-negative integers $m_1,m_2$\  and $m_3$,
where $m_{1}+m_{2}+2m_{3}=N$, and $m_1=0$\ or
$m_{2}=0$ must be satisfied in the case $m_3=0$.
\hfil\break
\indent
{}From explicit expression of transfer matrices at small $N$,
we obtain $\Lambda=a+b \pm (e+k)$ for $N=1$, and
$\Lambda=(a+b+e+k)^2,\ (a+b-e-k)^2,\ \pm (a-b+e-k)(a-b-e+k)$, for N=2
, which agree with the above formula. \vspace{3mm}\\
%%%%%%%%%%%%%%%%%%%%%%%%%%%%%%%%%%%%%%%%%%%%%%%%%%%%%%%%%%%%%%%
{\it ii)\ $ a=d, b=c, e=l, k=h$ case }\hfil\break
\indent
In this case, the Yang-Baxter equations are not satisfied but transfer
matrix commute. We first explain why the Yang-Baxter equations are not
satisfied in this case through the 8-vertex case,
because the expression becomes rather complicated in the 16-vertex case,
but the mechanism is the same.
Then we consider this special 8-vertex case, that is, $a=d,b=c,e=h=k=l=0$\
and the explicit Yang-Baxter equations give
\begin{eqnarray}
&& a b^{'} a^{''}+a a^{'} a^{''} =a b^{'} a^{''}+b a^{'} b{''}
\label{e86} \\
&& a b^{'} b^{''}+a a^{'} b^{''} =b a^{'} b^{''}+b b^{'} b{''}
\label{e87} \\
&&...   \nonumber
\end{eqnarray}
{}From Eqs.(\ref{e86}),(\ref{e87}), we obtain
$(a^{'}+b^{'})(a^{''}-b^{''})=0$
, but  as $a^{'}+b^{'} \ne 0$ because
Boltzmann weights must be positive, and as it is in general
$a^{''} \ne b^{''}$,
the Yang-Baxter equations are not satisfied in general.
Same mechanism happens in this 16-vertex case, and as it is in general
$a+d \ne b+c,\ e+l \ne k+h $, the Yang-Baxter equations
are not satisfied in this 16-vertex case.
\hfil \break
\indent
In this 16-vertex case, Eqs.(\ref{e43}), (\ref{e45}),
(\ref{e47}), (\ref{e49}) give
\begin{eqnarray}\left\{ \begin{array}{l}
(r_{i}^2-1)(r_{i+1}^2-1)=0,          \\
({s_{i}^{'}}^2-1)(s_{i}^2-1)=0,      \\
(r_{i}^2-1)(s_{i}^2-1)=0,            \\
(r_{i+1}^2-1)({s_{i}^{'}}^2-1)=0 \end{array} \right.  \label{e88}
\end{eqnarray}
Then we substitute solutions of Eq.(\ref{e88})
into the original pair propagation
and conjugate pair propagation equations, and we obtain following
cases ;\hfil\break
${iia)\ r_{i}=r_{i+1}=-s_{i}=-s_{i}^{'}=\pm 1\ case}$
\hfil\break
\indent
Choosing $p_{i}= p_{i+1}$\ , eigenfunctions at this site becomes
\begin{eqnarray}
g_{i}=
\left(\begin{array}{c}
g_{i}(+)  \\  \mp g_{i}(+) \\ \end{array}\right)
,\quad
g_{i}^{'}=0
,\quad
g_{i}^{''}=(a+b\mp (e+k)) g_{i}                               \label{e92}
\end{eqnarray}
${iib)\ r_{i}=-r_{i+1}=s_{i}=-s_{i}^{'}=\pm 1\ case}$
\hfil\break
\indent
In this case, choosing $p_{i}= p_{i+1}$\ , eigenfunctions
at this site become
\begin{eqnarray}
g_{i}=
\left(\begin{array}{c}
g_{i}(+)  \\ \pm g_{i}(+) \\ \end{array}\right)
,\quad
g_{i}^{'}=0
,\quad
g_{i}^{''}=(-a+b\pm (e-k))
\left(\begin{array}{c}
g_{i}(+)  \\  \mp g_{i}(+) \\ \end{array}\right)   \label{e94}
\end{eqnarray}
If we mix these cases, we obtain exactly the same
formula Eq.(\ref{e85}).
{}From the explicit expression of the transfer matrix of small $N$,
we obtain $\Lambda=a+b \pm (e+k)$ for $N=1$ and
$\Lambda=(a+b+e+k)^2,\ (a+b-e-k)^2,\ \pm(a-b+e-k)(a-b-e+k)$\ for $N=2$,
which agree with this formula. \vspace{10mm}
\hfil\break
%%%%%%%%%%%%%%%%%%%%%%%%%%% section 5 %%%%%%%%%%%%%%%%%%%%%%
\noindent {\bf 5. Summary and Discussion}\vspace{3mm}\\
\indent
We have clarified the connection between the Yang-Baxter and
the pair propagation equations in the 16-vertex models. In the
16-vertex models, we find exactly
solvable example of {\it i) $a=c,b=d,e=h,k=l$ } case,
where the Yang-Baxter equations are satisfied.
\hfil\break
\indent
We find another exactly solvable example of
{\it ii)\ $a=d,b=c,e=l,k=h$} case.
By explicit calculation, we can find that
conditions $ a+d=b+c,e+l=k+h$\ are necessary
to satisfy the Yang-Baxter equations in this case, but these are not
satisfied in general, that is, the Yang-Baxter equations
are not necessary condition for the solvability.
Though the Yang-Baxter equations
are not satisfied, we can show that transfer matrices
commute (integrable) for any lattice size $N$, which will be
discussed in a separate paper.
\hfil \break
\indent
In the 16-vertex models, from these exactly solvable examples,
integrable cases which satisfies the Yang-Baxter
equations are rather limited cases in the whole
exactly solvable cases.
In this sense, the pair propagation equations are more
fundamental, and even if the Yang-Baxter equations are not satisfied,
if the pair propagation equations are solvable, it is sufficient for our
purpose to find eigenvalues of tansfer matrices.\vspace{10mm}\\
%\newpage
%%%%%%%%%%%%%%%%%%%Acknowledgement %%%%%%%%%%%%%%%
\noindent {\bf Acknowledgement}\vspace{2mm}\\
\indent
One of the authors (K.S.) is grateful to the Special Research Fund at
Tezukayama Univ. for financial support. \vspace{10mm}
%\vfil \break
%\newpage
%%%%%%%%%%%%%%%%%%%%%%%%% references %%%%%%%%%%%%%%%%%%%

%%%%%%%%%%%%%%%%%%%%%%%%%%%%%%%%%%%%%%%%%%%%%%%%%%%%%%%%%%%%%%%
\end{document}